\documentclass[twocolumn,showpacs,preprintnumbers,amsmath,amssymb,superscriptaddress,prl]{revtex4}
\usepackage{epsfig}
\usepackage{graphicx}
\usepackage{color}
\usepackage{bm}

\begin{document}

\title{Detection of stacking faults breaking the [110]/[1$\bar{1}$0] symmetry in ferromagnetic semiconductors (Ga,Mn)As and (Ga,Mn)(As,P)}
\author{M.~Kopeck\'y}
\author{J.~Kub}
\author{F.~M\'aca}
\author{J.~Ma\v{s}ek}
\author{O.~Pacherov\'a}
\affiliation{Institute of Physics ASCR, v.v.i., Na Slovance 2, 182 21 Praha
8, Czech Republic}
\author{B.~L.~Gallagher}
\author{R.~P.~Campion}
\affiliation{School of Physics and Astronomy, University of Nottingham, Nottingham NG7 2RD, United Kingdom}
\author{V.~Nov\'ak}
\affiliation{Institute of Physics ASCR, v.v.i., Cukrovarnick\'a 10, 162 53 Praha
6, Czech Republic}
\author{T.~Jungwirth}
\affiliation{Institute of Physics ASCR, v.v.i., Cukrovarnick\'a 10, 162 53 Praha
6, Czech Republic} \affiliation{School of Physics and Astronomy,
University of Nottingham,
  Nottingham NG7 2RD, United Kingdom}

\begin{abstract}
We report high resolution x-ray diffraction  measurements of (Ga,Mn)As and (Ga,Mn)(As,P) epilayers. We observe a structural anisotropy in the form of stacking faults which are present in the (111) and (11$\bar{1}$)  planes and absent in the ($\bar{1}$11) and (1$\bar{1}$1) planes. The stacking faults produce no macroscopic  strain. They occupy $10^{-2}-10^{-1}$ per cent of the epilayer  volume. Full-potential density functional calculations evidence an attraction of Mn$_{\rm Ga}$ impurities to the stacking faults. We argue that  the enhanced Mn density along the common [1$\bar{1}$0] direction of the stacking fault planes produces sufficiently strong [110]/[1$\bar{1}$0] symmetry breaking mechanism to account for the in-plane uniaxial magnetocrystalline anisotropy of these ferromagnetic semiconductors. 
\end{abstract}


\maketitle

The rich phenomenology of magnetocrystalline anisotropies in ferromagnetic (III,Mn)V semiconductors has been the prerequisite of numerous studies of magnetic, magneto-transport, and magneto-optical phenomena and of prototype semiconductor spintronic devices \cite{Dietl:2008_b}. The in-plane biaxial anisotropy as well as the out-of-plane uniaxial anisotropy terms are well understood based on the periodic crystal structure characteristics of the epilayers \cite{Zemen:2009_a}. The former term is due to the cubic symmetry of the host III-V semiconductor and the latter term due to the lattice-matching strain induced by the difference between lattice parameters of the free-standing (III,Mn)V crystal and the substrate. 

The most extensively studied material is (Ga,Mn)As grown on GaAs, in which the compressive strain in the ferromagnetic (Ga,Mn)As epilayer makes the out-of-plane orientation of magnetization energetically unfavorable. It has been recognized from the early studies of these in-plane ferromagnets that the biaxial anisotropy term is complemented by an additional uniaxial term breaking the symmetry between [110] and [1$\bar{1}$0] crystal directions. While the presence of this term and its competition with the biaxial anisotropy term have played a key role in the research of ferromagnetic semiconductors, including studies of electrical or optical manipulation of the magnetic state \cite{Dietl:2008_b}, the microscopic origin of the in-plane uniaxial anisotropy has remained elusive. 

We have considered the following guidelines when searching for the uniaxial  in-plane symmetry breaking mechanism: (i) The corresponding uniaxial magnetocrystalline anisotropy is not a surface or interface effect but a bulk phenomenon. (ii) No strain component has been detected in the epilayers that would break the symmetry between the two in-plane diagonals. (iii) No systematic  dependence has been identified in the in-plane uniaxial magnetic anisotropy on the growth induced lattice-matching strain, as seen e.g. from the comparison of compressively strained (Ga,Mn)As/GaAs epilayers and tensile strained (Ga,Mn)(As,P)/GaAs epilayers. 
(iv) From the effective modeling of the in-plane uniaxial anisotropy it has been concluded that the symmetry breaking mechanism is related to the high Mn$_{\rm Ga}$ doping \cite{Zemen:2009_a}.

In this paper we report experimental observation and theoretical investigation of stacking faults in (Ga,Mn)As and (Ga,Mn)(As,P) which break the in-plane [110]/[1$\bar{1}$0] symmetry and whose characteristics are consistent with the above guidelines. The experimentally estimated density of the stacking faults and the theoretically inferred attraction of Mn$_{\rm Ga}$ to these lattice defects yields a strength  and sense of the symmetry breaking mechanism which we compare to the broken crystal symmetry due to an in-plane uniaxial strain. The latter symmetry breaking mechanism has been commonly used as an effective parameter to theoretically model the uniaxial magnetocrystalline energy in unpatterned epilayers or as a real tool to control the anisotropy in microstructured films  or in epilayers attached to piezostressors \cite{Zemen:2009_a}. 

Measured (Ga,Mn)As and (Ga,Mn)(As,P) samples were grown by  low-temperature (200-230$^\circ$C) molecular beam epitaxy (LT-MBE) on a GaAs substrate and buffer layer. For more details on the sample growth see Refs.~\cite{Jungwirth:2010_b,Rushforth:2008_b}. X-ray experiments were carried out on the diffraction beamline at the ELETTRA synchrotron facility in Trieste. Samples were mounted on a two-axis tilt platform and aligned in the way that the normal to the sample surface coincides with the rotation axis  of the diffractometer; the rotation angle is $\phi$. In this geometry, the glancing angle $\theta$ between the beam and the sample surface remained constant during the $\phi$-scan. The energy of the incident beam was set to 10.3~keV, i.e. just below the $K$ absorption edge of gallium, in order to minimize the absorption in the sample and to avoid fluorescence. The beam size was set to 500~$\mu$m (horizontal) and 20~$\mu$m (vertical) using two pairs of slits in front of the sample. A small vertical aperture guarantees the elimination of the undesired scattering from sample borders. The area detector Pilatus 2M ($1475\times1679$ pixels of the size of $172\times172$~$\mu$m$^2$ in $3\times8$ modules, dynamic range of 20 bits) made it possible to collect the weak diffuse scattering patterns using the exposure rate of 5~s per frame. One scan was composed of 720 frames collected with the angular step $\Delta\phiÊ=Ê0.5^{\circ}$. The frames were put together and the intensity maps in the large volume of the reciprocal space were constructed. 

In Figs.~1(a),(b) we show measured data for  500~nm thick 5\% Mn-doped (Ga,Mn)As epilayer. The large sample thickness allows us to set the glancing angle $\theta$ to 0.3$^{\circ}$, which is slightly above the critical angle $\theta_cÊ=Ê0.25^{\circ}$. Under these conditions a large portion of the (Ga,Mn)As epilayer volume is illuminated while only a small portion of the beam penetrates into the substrate and the scattering from the substrate is negligible. Two cross-sections through the reciprocal space perpendicular to each other are shown in Figs.~1(a),(b). The cross-section in panel 1(a) contains only diffraction spots along with truncation rods perpendicular to the sample surface. On the other hand, the cross-section shown in panel 1(b) contains additional diffuse streaks in directions [111] and [11$\bar{1}$], indicating the presence of stacking faults in planes (111) and (11$\bar{1}$). 

In Fig.~2(a) we present a detailed image of the area near the -111 diffraction and compare in Fig.~2(b) with the same area measured at $\theta=0.2^{\circ}$, i.e.,  slightly below the critical angle. In the latter image the thickness of the illuminated (Ga,Mn)As surface layer is only $\sim 10$~nm. Consistently, the truncation rod perpendicular to the sample surface is more extended in Fig.~2(b) than 2(a). The comparison of the two images demonstrates that the stacking faults occupy a significant part of the epilayer volume since the intensity of the diffuse [111] and [11$\bar{1}$] streaks  is significantly larger in Fig.~2(a) than 2(b). 

We have performed additional measurements at $\theta=0.2^{\circ}$ on (Ga,Mn)As and (Ga,Mn)(As,P) epilayers of width ranging from 35 to 100~nm, as-grown and annealed, and with Mn-doping up to 10\% and P-doping of 9\%. In all cases we obtained results similar to Fig.~2(b) with the diffuse streaks only in directions [111] and [11$\bar{1}$]. No diffuse streaks in any direction were observed in a reference, LT-MBE grown undoped 500~nm thick GaAs epilayers.  From these data we can conclude that the stacking faults span across the epilayer volume, are related to the Mn-doping, are present in the same planes in samples with smaller (annealed GaMnAs) and larger (as-grown GaMnAs) magnitude of the growth lattice-matching strain and in epilayers with both compressive (GaMnAs) and tensile (GaMnAsP) strain. We also point out that the stacking faults are present in thick epilayers as well as in thin, annealed high magnetic quality films.

\begin{figure}[ht]
\vspace*{-0.cm}
\hspace*{-0.cm}\includegraphics[height=1.6\columnwidth,angle=0]{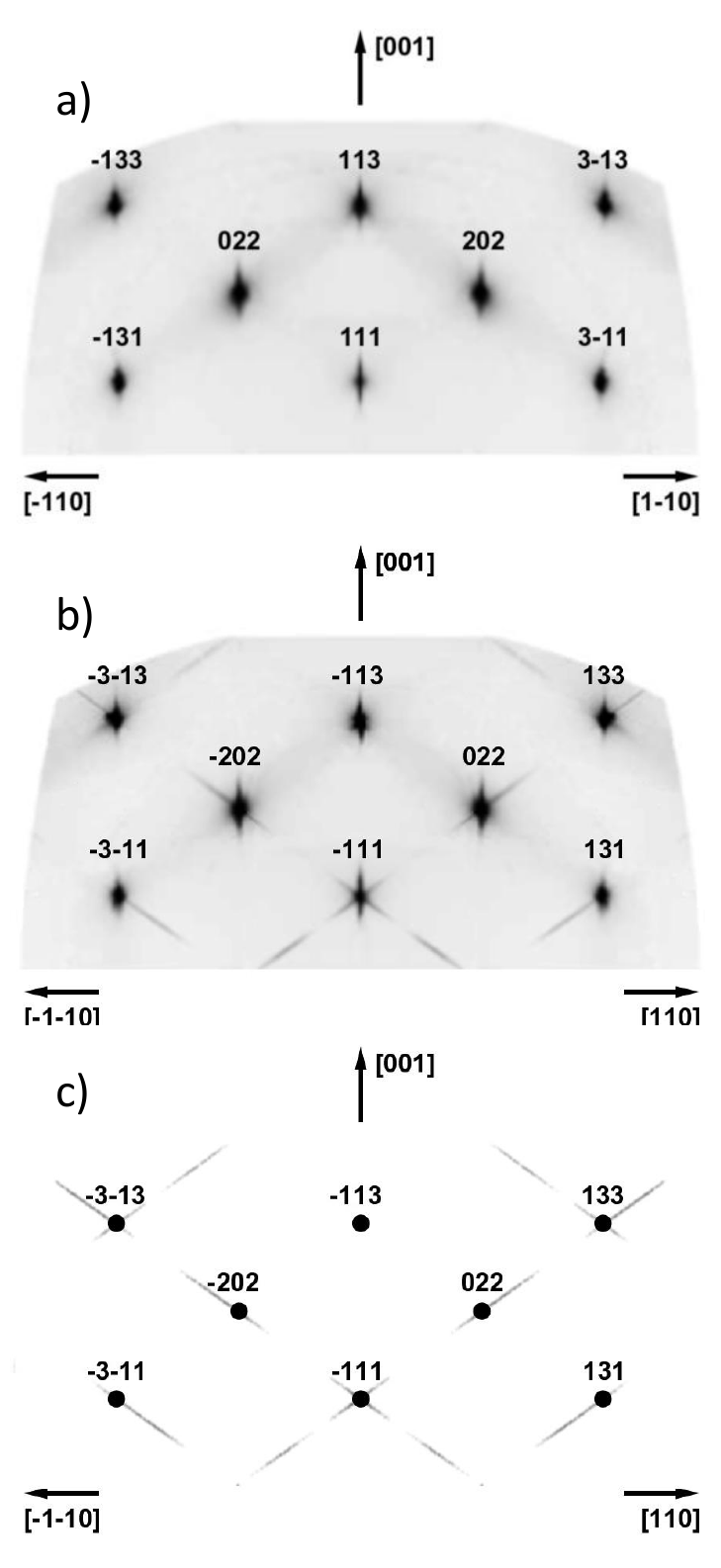}
\vspace*{-.5cm} \caption{(a,b)ÊÊCuts through the reciprocal space of the measured diffraction intensity maps. The vertical  truncation rods perpendicular to the sample surface are due to finite thickness of the illuminated film. The streaks in directions [111] and [11-1] represent the diffraction on stacking faults in planes (111) and (11-1), resp. (c)ÊÊThe calculated x-ray diffraction considering the distribution of the stacking faults described in the text for the cut through the reciprocal space as in (b).}
\label{Figure1}
\end{figure}

\begin{figure}[ht]
\vspace*{-0.cm}
\hspace*{-0.cm}\includegraphics[height=.45\columnwidth,angle=0]{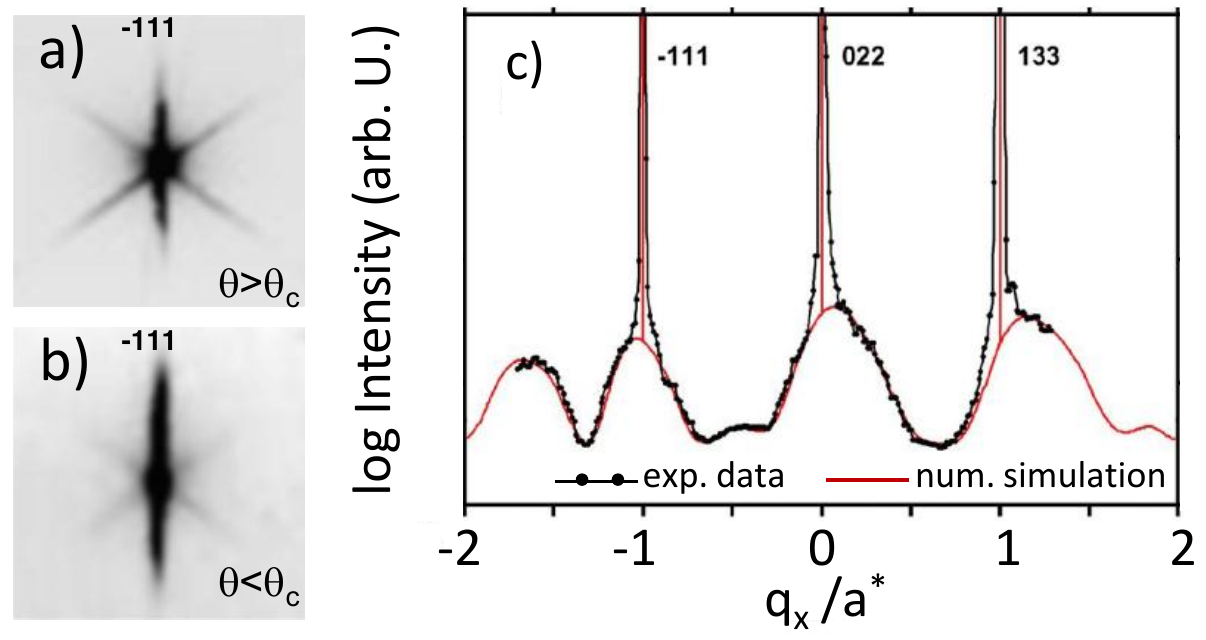}
\vspace*{-0cm} \caption{Detail of the diffuse scattering in the vicinity of the reflection -111 measured at (a) $\theta=Ê0.3^{\circ}$ and  (b) $\theta=Ê0.2^{\circ}$. (c) Measured intensity profile along the [111] direction passing through the reflection -111 and corresponding numerical simulation. Here $q_x$ is the $x$ co-ordinate of the diffraction vector and $a^\ast$ is the reciprocal lattice unit cell vector.}
\label{Figure2}
\end{figure}
From the modeling of the measured x-ray scattering patters (Figs.~1(c) and 2(c)), we identified the microscopic nature of the stacking faults. In the unperturbed GaAs zinc-blende crystal structure, the plane stacking along the [111] direction (or any other body diagonal) is characterized by a repeating a-b-c sequence of Ga-As planes, as shown in Fig.~3(a). A stacking fault along the [111] diagonal  shown in Fig.~3(b) corresponds to one of the planes missing in the stacking. The fault can be also viewed as replacing the zinc-blende stacking by two interpenetrating wurtzite stackings highlighted by the two red (vertical) bars in Fig.~3(b). A stacking fault due to an extra plane, or more separated two wurtzite sequences, is depicted in Fig.~3(c). A removed and immediately inserted plane creates a fault comprising two neighboring wurtzite stackings, as shown in Fig.~3(d). Our experimental data could not be fitted  by the stacking faults in 3(b) and 3(c). As seen from the comparison with Fig.~3(a), these two faults yield a zinc-blende stacking after the fault which is shifted by one plane with respect to the unperturbed lattice. It disrupts the matching of the epilayer with these stacking faults to the substrate on a macroscopic scale and, therefore, the faults are unlikely to form. Their absence is also consistent with the absence of an in-plane uniaxial strain which we confirmed experimentally in our epilayers with accuracy $\sim 10^{-5}$. (We compared average lattice parameters in the [110] and [1$\bar{1}$0] directions by using relative position of 17 Bragg reflections of the Mn-doped epilayer with respect to these of the GaAs substrate.) The stacking fault in Fig.~3(d), on the other hand, produces no macroscopic strain and indeed our fittings show that this fault represents $\sim 90$\% of the stacking faults present in the measured epilayers. The remaining faults, producing also no macroscopic strain outside the fault, have one or more zinc-blende a-b-c sequences between the removed and inserted plane and their probability with increasing number of inner a-b-c sequences was fitted by a $\Gamma$-function distribution. 

\begin{figure}[ht]
\vspace*{-0.cm}
\hspace*{-0.cm}\includegraphics[height=1.15\columnwidth,angle=0]{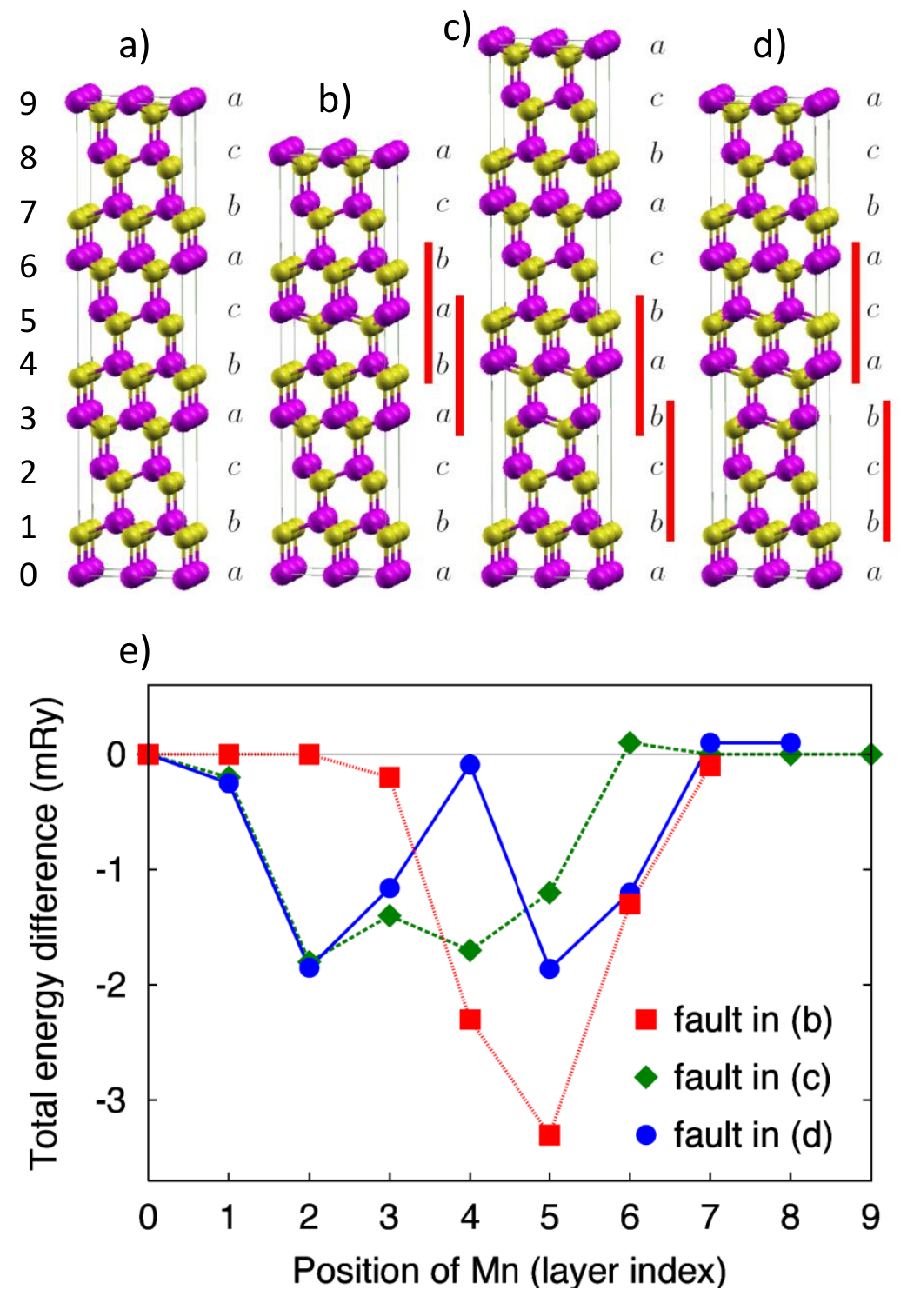}
\vspace*{-0.2cm} \caption{(a) Zinc-blende stacking in an ideal GaAs lattice. (b)-(d) Stacking faults with one removed, added, or removed and immediately inserted planes, resp. (e) Calculations of the total energy of the supercell with the respective stacking fault as a function of the position of the substitutional Mn$_{\rm Ga}$. Total energies are plotted with respect to the energy of the supercell with Mn$_{\rm Ga}$ at the 0-th layer in the supercell.}
\label{Figure3}
\end{figure}

As pointed out in the previous paragraph, there is no experimentally observable strain in the (Ga,Mn)As epilayers, within the $\sim 10^{-5}$ error bar which would break the [110]/[1$\bar{1}$0] crystal symmetry. The uniaxial  in-plane  strain has been used as a hypothetical symmetry breaking mechanism to account in microscopic theory calculations for the experimentally observed uniaxial in-plane magneto-crystalline anisotropy \cite{Sawicki:2004_a,Zemen:2009_a}. Typical magnitudes of the strain required by the modeling to reproduce the measured magnitude of the magnetic anisotropy are $\sim10^{-4}$, i.e., an order of magnitude larger  than the experimental error of our measurements. Based on this we can exclude strain as the real microstructural origin of the [110]/[1$\bar{1}$0] magnetic anisotropy. Results of our {\em ab initio} calculations, described in the following paragraphs, suggest that the observed stacking faults provide the searched for microstructural symmetry breaking mechanism. 
\begin{figure}[ht]
\vspace*{-0.cm}
\hspace*{-0.cm}\includegraphics[height=0.6\columnwidth,angle=0]{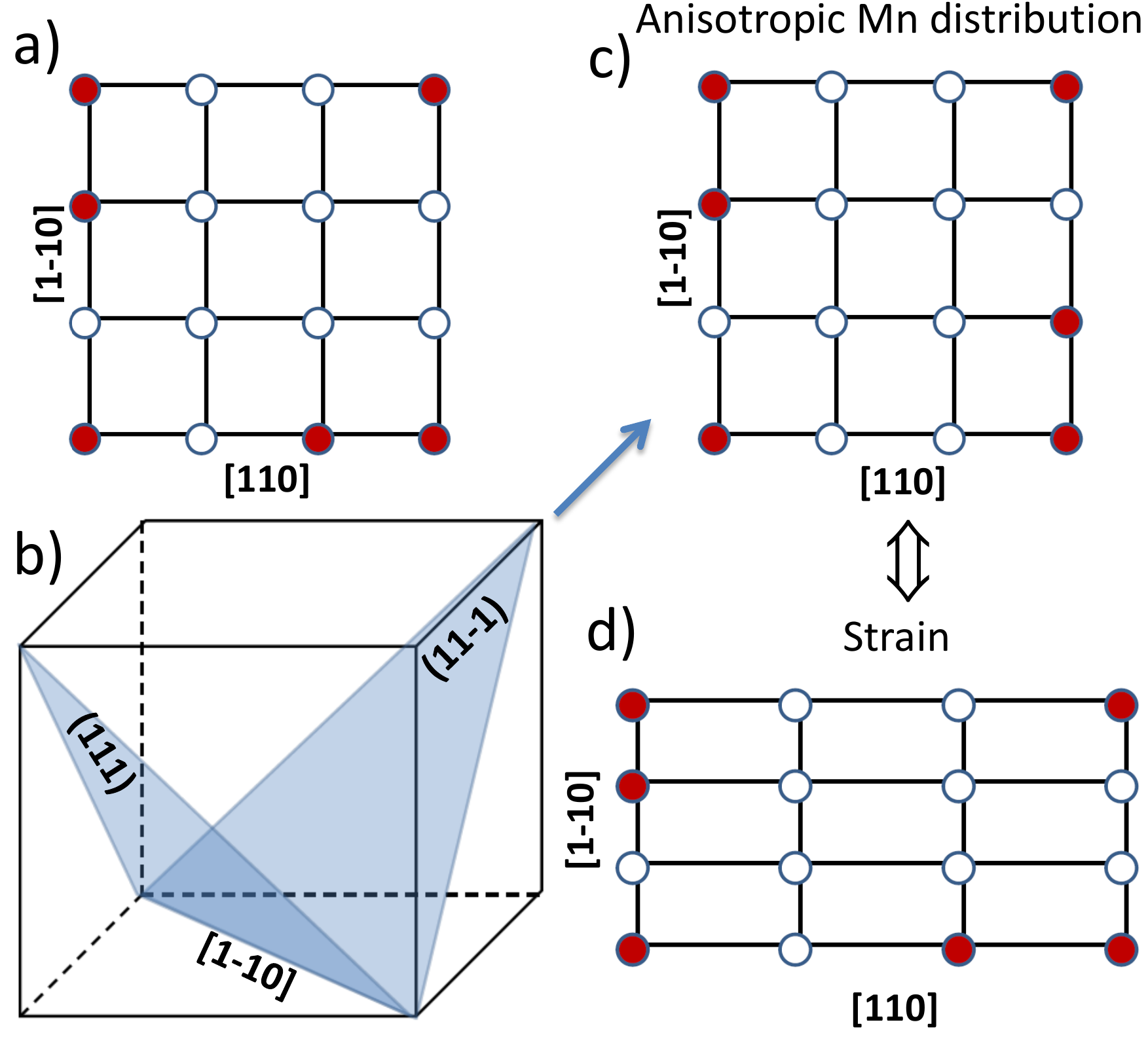}
\vspace*{-0.cm} \caption{(a) Schematic image of uniformly randomly distributed Mn (red (dark) dots) in the epilayer plane in an unstrained lattice with cubic symmetry. (b) Schematics of the planes in which the stacking faults are observed  by x-ray diffraction. (c) Schematic image of anisotropic Mn distribution in the epilayer plane in an unstrained lattice which results from the tendency of Mn to decorate the stacking faults. (d) Schematic image of uniformly distributed Mn on the lattice sites in the presence of a uniaxial in-plane symmetry breaking strain. (c) and (d) are analogous from the perspective of the anisotropic mean separation of Mn in the epilayer plane.}
\label{Figure4}
\end{figure}
Our supercell calculations were performed using the full-potential linearized-augmented-plane-wave method with the improved local-density approximation by the generalized-gradient approximation (LAPW-GGA WIEN package \cite{WIEN97}). The  calculated formation energies of the stacking faults shown in Figs.~3(b)-(d) for an undoped GaAs are in the range of 13-15~mRy. In Fig.~3(e) we plot the dependence of the formation energy of the stacking faults on the position of the substitutional Mn$_{\rm Ga}$ impurity along the [111] (or [11$\bar{1}$]) crystal direction in the supercell. By comparing the results for the three considered  stacking faults (Figs.~3(b)-(d)) we can conclude that  Mn$_{\rm Ga}$ is attracted to the wurtzite stacking perturbation in the lattice and the energy profile can be understood as a convolution of attractive potentials due to the individual wurtzite perturbations. Remarkably, in our supercells corresponding to 25\% of Mn$_{\rm Ga}$ impurities, the depth of the attractive potential energy well is as large as 20\% of the formation energy of the stacking fault in GaAs. We conclude that Mn impurities have a strong tendency to decorate the stacking faults and that by this process the Mn impurities significantly enhance the probability of the formation of the stacking faults in the epilayer.

In schematic diagrams in Fig.~4 we illustrate the inferred connection between the Mn$_{\rm Ga}$ decorated stacking faults and the uniaxial magnetocrystalline anisotropy. Fig.~4(a) shows schematically the biaxial symmetry in case of unperturbed lattice and uniform distribution of Mn$_{\rm Ga}$ impurities. Since the stacking faults are observed only in the (111) and (11$\bar{1}$) planes (Fig.~4(b)) and Mn$_{\rm Ga}$ is attracted to the stacking faults, we can expect higher mean density of the Mn$_{\rm Ga}$ impurities along the common [1$\bar{1}$0] direction of the two planes. This leads to an anisotropic distribution of  Mn$_{\rm Ga}$ impurities in the epilayer plane with a smaller mean distance between impurities in the [1$\bar{1}$0] direction than [110] direction.

In Fig.~4(d) we illustrate that the non-uniform Mn$_{\rm Ga}$ distribution in an unstrained square lattice  is from the perspective of the mean Mn$_{\rm Ga}$ separation analogous to the case of a uniform Mn distribution and a strained lattice. The latter case has been considered previously in the effective modelings of the corresponding uniaxial magnetic anisotropy. Since the density of the stacking faults we observe is $\sim 0.01-0.1$\%, the corresponding symmetry breaking mechanism is strong enough to explain the uniaxial magnetic anisotropy.  (Compare the density of the stacking faults with the $\sim10^{-4}$ strain used in the effective modeling of the magnetic anisotropy.) We note that detection of the stacking faults and of the increased Mn$_{\rm Ga}$ density on these defects is likely unachievable by direct atomic resolution scanning techniques. This underlines the merit of our approach, combining the macro-scale x-ray measurements with {\em ab initio} total energy calculations, for identifying the subtle symmetry breaking mechanism. 

The sign of the magnetic anisotropy in the Mn-doped magnetic epilayers is not uniquely determined by the sign of the underlying broken structural symmetry; it can vary for the given sense of the broken symmetry as a function of, e.g., the hole density in the magnetic film. Typically, however, the [1$\bar{1}$0] direction is magnetically easier than [110] direction. The common [1$\bar{1}$0] direction of the stacking fault planes then coincides with the typical easy magnetic direction. If again considering the analogy with the case of uniform Mn$_{\rm Ga}$ distribution and in-plane strained lattice, this observation provides further support for the stacking-fault origin of the uniaxial in-plane anisotropy. From the modelings considering the in-plane strain, as well as from experiments on intentionally in-plane strained (Ga,Mn)As epilayers by lithographical pattering or piezostressors \cite{Zemen:2009_a}, it has been concluded that the magnetic easy direction is along the smaller lattice parameter, i.e., along the direction with  smaller mean separation between Mn$_{\rm Ga}$ impurities. Our results show that in unstrained epilayers the stacking faults yield smaller mean separation between Mn$_{\rm Ga}$ impurities along the in-plane axis which also coincides with the typical  uniaxial magnetic easy direction. We conclude that all characteristics of the experimentally observed stacking faults in the studied (III,Mn)V epilayers are consistent with the guidelines (i)-(iv) summarized in the introduction for identifying the microstructural [110]/[1$\bar{1}$0] symmetry breaking mechanism in these magnetic semiconductors.

We acknowledge support from EU Grants FP7-215368 SemiSpinNet, FP7-214499 NAMASTE, and Czech Republic Grants  GAAV IAA100100915, GACR 202/07/0456, AV0Z10100520, AV0Z10100521, KAN400100652, LC510, Preamium Academiae.


\begin{thebibliography}{6}
\expandafter\ifx\csname natexlab\endcsname\relax\def\natexlab#1{#1}\fi
\expandafter\ifx\csname bibnamefont\endcsname\relax
  \def\bibnamefont#1{#1}\fi
\expandafter\ifx\csname bibfnamefont\endcsname\relax
  \def\bibfnamefont#1{#1}\fi
\expandafter\ifx\csname citenamefont\endcsname\relax
  \def\citenamefont#1{#1}\fi
\expandafter\ifx\csname url\endcsname\relax
  \def\url#1{\texttt{#1}}\fi
\expandafter\ifx\csname urlprefix\endcsname\relax\def\urlprefix{URL }\fi
\providecommand{\bibinfo}[2]{#2}
\providecommand{\eprint}[2][]{\url{#2}}

\bibitem[{\citenamefont{eds. T.~Dietl et~al.}(2008)\citenamefont{eds. T.~Dietl,
  Awschalom, Kaminska, and Ohmo}}]{Dietl:2008_b}
\bibinfo{author}{\bibnamefont{eds. T.~Dietl}},
  \bibinfo{author}{\bibfnamefont{D.~D.} \bibnamefont{Awschalom}},
  \bibinfo{author}{\bibfnamefont{M.}~\bibnamefont{Kaminska}}, \bibnamefont{and}
  \bibinfo{author}{\bibfnamefont{H.}~\bibnamefont{Ohmo}}, in
  \emph{\bibinfo{booktitle}{Spintronics}} (\bibinfo{publisher}{Elsevier},
  \bibinfo{year}{2008}), vol.~\bibinfo{volume}{82} of
  \emph{\bibinfo{series}{Semiconductors and Semimetals}}.

\bibitem[{\citenamefont{Zemen et~al.}(2009)\citenamefont{Zemen, Kucera,
  Olejnik, and Jungwirth}}]{Zemen:2009_a}
\bibinfo{author}{\bibfnamefont{J.}~\bibnamefont{Zemen}},
  \bibinfo{author}{\bibfnamefont{J.}~\bibnamefont{Kucera}},
  \bibinfo{author}{\bibfnamefont{K.}~\bibnamefont{Olejnik}}, \bibnamefont{and}
  \bibinfo{author}{\bibfnamefont{T.}~\bibnamefont{Jungwirth}},
  \bibinfo{journal}{Phys. Rev.} \textbf{\bibinfo{volume}{B 80}},
  \bibinfo{pages}{155203} (\bibinfo{year}{2009}).

\bibitem[{\citenamefont{Jungwirth et~al.}(2010)\citenamefont{Jungwirth,
  {Horodysk\'{a}}, {Tesa\v{r}ov\'{a}}, {N\v{e}mec}, {\v{S}ubrt}, {Mal\'{y}},
  {Ku\v{z}el}, Kadlec, {Ma\v{s}ek}, {N\v{e}mec} et~al.}}]{Jungwirth:2010_b}
\bibinfo{author}{\bibfnamefont{T.}~\bibnamefont{Jungwirth}},
  \bibinfo{author}{\bibfnamefont{P.}~\bibnamefont{{Horodysk\'{a}}}},
  \bibinfo{author}{\bibfnamefont{N.}~\bibnamefont{{Tesa\v{r}ov\'{a}}}},
  \bibinfo{author}{\bibfnamefont{P.}~\bibnamefont{{N\v{e}mec}}},
  \bibinfo{author}{\bibfnamefont{J.}~\bibnamefont{{\v{S}ubrt}}},
  \bibinfo{author}{\bibfnamefont{P.}~\bibnamefont{{Mal\'{y}}}},
  \bibinfo{author}{\bibfnamefont{P.}~\bibnamefont{{Ku\v{z}el}}},
  \bibinfo{author}{\bibfnamefont{C.}~\bibnamefont{Kadlec}},
  \bibinfo{author}{\bibfnamefont{J.}~\bibnamefont{{Ma\v{s}ek}}},
  \bibinfo{author}{\bibfnamefont{I.}~\bibnamefont{{N\v{e}mec}}},
  \bibnamefont{et~al.}, \bibinfo{journal}{Phys. Rev. Lett.}
  \textbf{\bibinfo{volume}{105}}, \bibinfo{pages}{227201}
  (\bibinfo{year}{2010}).

\bibitem[{\citenamefont{Rushforth et~al.}(2008)\citenamefont{Rushforth, Wang,
  Farley, Campion, Edmonds, Staddon, Foxon, and Gallagher}}]{Rushforth:2008_b}
\bibinfo{author}{\bibfnamefont{A.~W.} \bibnamefont{Rushforth}},
  \bibinfo{author}{\bibfnamefont{M.}~\bibnamefont{Wang}},
  \bibinfo{author}{\bibfnamefont{N.~R.~S.} \bibnamefont{Farley}},
  \bibinfo{author}{\bibfnamefont{R.~C.} \bibnamefont{Campion}},
  \bibinfo{author}{\bibfnamefont{K.~W.} \bibnamefont{Edmonds}},
  \bibinfo{author}{\bibfnamefont{C.~R.} \bibnamefont{Staddon}},
  \bibinfo{author}{\bibfnamefont{C.~T.} \bibnamefont{Foxon}}, \bibnamefont{and}
  \bibinfo{author}{\bibfnamefont{B.~L.} \bibnamefont{Gallagher}},
  \bibinfo{journal}{J. Appl. Phys.} \textbf{\bibinfo{volume}{104}},
  \bibinfo{pages}{073908} (\bibinfo{year}{2008}).

\bibitem[{WIE()}]{WIEN97}
\bibinfo{note}{P.~Blaha, K.~Schwarz, G. K.H. Madsen, D. Kvasnicka, and J.~Luitz, WEIN2k, FPLAPW package for
  calculating crystal properties, TU Vienna.}

\bibitem[{\citenamefont{Sawicki et~al.}(2005)\citenamefont{Sawicki, Wang,
  Edmonds, Campion, Staddon, Farley, Foxon, Papis, Kaminska, Piotrowska
  et~al.}}]{Sawicki:2004_a}
\bibinfo{author}{\bibfnamefont{M.}~\bibnamefont{Sawicki}},
  \bibinfo{author}{\bibfnamefont{K.-Y.} \bibnamefont{Wang}},
  \bibinfo{author}{\bibfnamefont{K.~W.} \bibnamefont{Edmonds}},
  \bibinfo{author}{\bibfnamefont{R.~P.} \bibnamefont{Campion}},
  \bibinfo{author}{\bibfnamefont{C.~R.} \bibnamefont{Staddon}},
  \bibinfo{author}{\bibfnamefont{N.~R.~S.} \bibnamefont{Farley}},
  \bibinfo{author}{\bibfnamefont{C.~T.} \bibnamefont{Foxon}},
  \bibinfo{author}{\bibfnamefont{E.}~\bibnamefont{Papis}},
  \bibinfo{author}{\bibfnamefont{E.}~\bibnamefont{Kaminska}},
  \bibinfo{author}{\bibfnamefont{A.}~\bibnamefont{Piotrowska}},
  \bibnamefont{et~al.}, \bibinfo{journal}{Phys. Rev.}
  \textbf{\bibinfo{volume}{B 71}}, \bibinfo{pages}{121302}
  (\bibinfo{year}{2005}).

\end{thebibliography}
\end{document}